\documentclass[a4paper,preprint]{aastex}

\usepackage{natbib}

\begin{document}

\title{A Search for a Sub-Earth Sized Companion to GJ 436 and a Novel Method to Calibrate Warm Spitzer IRAC Observations}

\author{Sarah~Ballard\altaffilmark{1}, David~Charbonneau\altaffilmark{1}, Drake~Deming\altaffilmark{2}, Heather~A.~Knutson\altaffilmark{3,4}, Jessie~L.~Christiansen\altaffilmark{5}, Matthew~J.~Holman\altaffilmark{1}, Daniel~Fabrycky\altaffilmark{1,6}, Sara Seager\altaffilmark{7}, Michael~F.~A'Hearn\altaffilmark{8}}

\altaffiltext{1}{Harvard-Smithsonian Center for Astrophysics, 60 Garden Street, Cambridge, MA 02138, USA; sballard@cfa.harvard.edu}
\altaffiltext{2}{Planetary Systems Branch, Code 693, NASA/Goddard Space Flight Center, Greenbelt, MD 20771, USA}
\altaffiltext{3}{Department of Astronomy, 601 Campbell Hall, University of California, Berkeley, CA 94720-3411, USA}
\altaffiltext{4}{Miller Fellow}
\altaffiltext{5}{NASA Ames Research Center, MS-244-30, P.O. Box 1, Moffett Field, CA 94035-0001, USA}
\altaffiltext{6}{Michelson Fellow}
\altaffiltext{7}{Department of Earth, Atmospheric, and Planetary Sciences, 77 Massachusetts Avenue, Massachusetts Institute of Technology, Cambridge, MA 02159, USA}
\altaffiltext{8}{Department of Astronomy, University of Maryland, College Park, MD 20742-2421, USA}

\begin{abstract}
We discovered evidence for a possible additional 0.75 $R_{\oplus}$ transiting planet in the NASA {\it EPOXI} observations of the known M~dwarf exoplanetary system GJ~436. Based on an ephemeris determined from the {\it EPOXI} data, we predicted a transit event in an extant {\it Spitzer Space Telescope} 8 $\mu$m data set of this star. Our subsequent analysis of those {\it Spitzer} data confirmed the signal of the predicted depth and at the predicted time, but we found that the transit depth was dependent on the aperture used to perform the photometry. Based on these suggestive findings, we gathered new Warm Spitzer Observations of GJ~436 at 4.5 $\mu$m spanning a time of transit predicted from the {\it EPOXI} and {\it Spitzer} 8 $\mu$m candidate events. The 4.5 $\mu$m data permit us to rule out a transit at high confidence, and we conclude that the earlier candidate transit signals resulted from correlated noise in the {\it EPOXI} and {\it Spitzer} 8 $\mu$m observations. In the course of this investigation, we developed a novel method for correcting the intrapixel sensitivity variations of the 3.6 and 4.5 $\mu$m channels of the Infrared Array Camera (IRAC) instrument. We demonstrate the sensitivity of Warm {\it Spitzer} observations of M~dwarfs to confirm sub-Earth sized planets. Our analysis will inform similar work that will be undertaken to use Warm {\it Spitzer} observations to confirm rocky planets discovered by the {\slshape Kepler} mission.
\end{abstract}


\section{Introduction}

With the recent discoveries of the transiting super Earths CoRoT-7b \citep{Leger09} and GJ~1214b \citep{Charbonneau09}, and the launch of the {\slshape Kepler} Mission \citep{Borucki10}, astronomers have begun to probe the regime of super-Earth exoplanets. CoRoT-7b, with a radius of 1.7 $R_{\oplus}$ in an orbit around a 0.87 $R_{\odot}$ star, produces a photometric signal of only 340 ppm \citep{Leger09}. The radial velocity confirmation of CoRoT-7b required 70 hours of follow-up observations with the HARPS instrument \citep{Queloz09}. A complementary approach is to use Warm {\it Spitzer} observations to prove that the transit depth is not color dependent \citep{Fressin10}. Similar follow-up observations using Warm {\it Spitzer} to confirm candidates identified by {\slshape Kepler} are already being gathered as part of an Exploration Science Program (D. Charbonneau). In this work, we present a search for a 0.75 $R_{\oplus}$ transiting planet around the M dwarf GJ~436, which is already known to host the transiting hot Neptune GJ~436b in an eccentric orbit \citep{Butler04,Maness07,Gillon07b, Deming07, Demory07}. 

 {\it EPOXI} is a NASA Discovery Program Mission of Opportunity using the Deep Impact flyby spacecraft \citep{Blume05}, comprising the Extrasolar Planet Observation and Characterization (EPOCh) investigation and the Deep Impact eXtended Investigation (DIXI). From January through August 2008, the EPOCh Science Investigation used the HRI camera \citep{Hampton05} with its broad visible bandpass to gather precise, rapid cadence photometric time series of known transiting exoplanet systems \citep{Ballard10, Christiansen09}. The majority of these targets were each observed nearly continuously for several weeks at a time. 

 One of the {\it EPOXI} science goals was a search for additional planets in these systems. Such planets would be revealed either through the variations they induce on the transit times of the known exoplanet, or directly through the transit of the second planet itself. This search is especially interesting in the case of the GJ~436 system. The eccentricity of the known transiting Neptune-mass planet, GJ~436b \citep{Butler04}, may indicate the presence of an additional perturbing planet, since the assumed circularization timescale for the known planet is much less than the age of the system \citep{Maness07, Deming07, Demory07}. \cite{Ribas08} claimed evidence for a 5 $M_{\oplus}$ super-Earth in radial velocity observations of GJ~436, but this proposed planet was ruled out by subsequent investigations \citep{Alonso08,Bean08b}. The absence of this additional perturbing body in the GJ~436 system would also be very scientifically interesting. If no other body is present to explain the eccentricity of GJ~436b, the observed eccentricity requires a very high tidal dissipation parameter, $Q$. 

We presented our search for additional transiting planets in the {\it EPOXI} observations of GJ~436 in \cite{Ballard10}. We demonstrated the sensitivity to detect additional transiting planets as small as 1.5 $R_{\oplus}$ interior to GJ~436b. We further uncovered evidence for a 0.75 $R_{\oplus}$ transiting planet, in a orbit close to a 4:5 resonance with GJ~436b, below the formal detection limit established by \cite{Ballard10}. We first analyzed an extant 8 $\mu$m {\it Spitzer Space Telescope} phase curve of GJ~436, obtained two months after the {\it EPOXI} observations as part of {\it Spitzer} Program 50056 (PI: H. Knutson). We then gathered Warm {\it Spitzer} 4.5 $\mu$m observations of GJ~436, which enabled us to conclusively test the planet hypothesis. We found that the current state-of-the-art reduction techniques to remove the intrapixel sensitivity variations associated with the Infrared Array Camera (IRAC) instrument \citep{Reach05, Charbonneau05, Knutson08, Knutson09} were insufficient to remove correlated noise at an amplitude comparable to the depth of the putative transit. We therefore pursued a novel technique for the removal of this intrapixel sensitivity variation. When compared to the earlier method, our technique identifies and corrects for high frequency intrapixel sensitivity features which were previously missed. Our novel method enhances the sensitivity of Warm {\it Spitzer} observations to transits of sub-Earth sized planets.

The remainder of this paper is organized as follows. In Section 2, we describe the observations and the photometry time series extraction for the {\it EPOXI} and {\it Spitzer} data sets. Section 2.3 describes the novel technique used to reduce the 4.5 $\mu$m observations. In Section 3, we consider the evidence for the planet hypothesis in the three data sets and we demonstrate the sensitivity of the Warm {\it Spitzer} 4.5 $\mu$m observations of GJ~436 to detect a 0.75 $R_{\oplus}$ planet. In Section 4, we discuss the applications of this work for future transit searches, including those to confirm candidate rocky planets from the {\slshape Kepler} mission.

\section{Observations and Time Series Extraction}
 
\subsection{{\it EPOXI} Observations}
\begin{figure}[h!]
\begin{center}
 \includegraphics[width=6in]{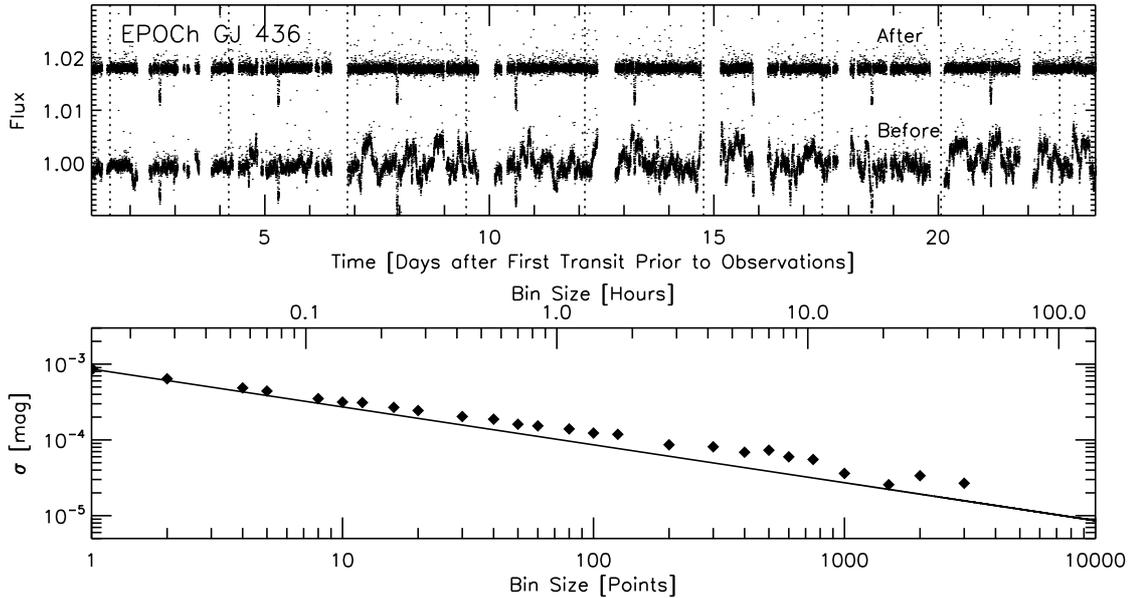} 
 \caption{\textit{Top panel:} GJ~436 time series before (lower curve) and after (upper curve) 2D spline correction. The uncorrected time series (lower curve) has had the two middle rows and columns in each image, and the entire image if observed in the large sub-array mode, scaled by a multiplicative factor to reduce the flux dependence on position and sub-array size. We have used the 2D spline to correct for additional interpixel variation in the upper curve. \textit{Bottom panel:} The data (diamond symbols) bin down consistently with the expectation for Gaussian noise (shown with a line, normalized to match the value at N=1).}
   \label{fig:lightcurve}
\end{center}
\end{figure}

We acquired observations of GJ~436 nearly continuously during 2008 May 5---29, interrupted for several hours at approximately 2-day intervals for data downloads. A complete description of the {\it EPOXI} photometric extraction pipeline is given in \cite{Ballard10} and summarized here. We used the existing Deep Impact data reduction pipeline to perform bias and dark subtractions, as well as preliminary flat fielding \citep{Klaasen05}. We first determined the position of the star on the CCD using PSF fitting, by maximizing the goodness-of-fit (with the $\chi^{2}$ statistic as an estimator) between an image and a model PSF (oversampled by a factor of 100) with variable position, additive sky background, and multiplicative brightness scale factor. We then processed the images to remove sources of systematic error due to the CCD readout electronics. We first scaled down the two central rows by a constant value, then we scaled down the central columns by a separate constant value, and finally we scaled the entire image by a multiplicative factor determined by the size of the sub-array. We performed aperture photometry on the corrected images, using an aperture radius of 10 pixels, corresponding to twice the HWHM of the PSF. To remove remaining correlated noise due to the interpixel sensitivity variations on the CCD, we fit a 2D spline surface, with the same resolution as the CCD, to the brightness variations on the array as follows. We randomly drew a subset of several thousand out-of-transit and out-of-eclipse points from the light curve (from a data set of $\sim$29,988 points) and found a robust mean of the brightness of the 30 nearest neighbors for each. We fit a spline surface to these samples and corrected each data point individually by linearly interpolating on this best-fit surface. We used only a small fraction of the observations to create the spline surface in order to minimize the potential transit signal suppression introduced by flat fielding the data with itself. To produce the final time series, we iterated the above steps, fitting for the row and column multiplicative factors, the sub-array size scaling factor, and the 2D spline surface that minimized the out-of-transit white noise of the photometric time series. We included one additional step to create the final 2D spline, which was to iteratively remove an overall modulation from the GJ~436 light curve which we attributed to star spots. After we took these steps to address the systematics associated with the observations, the red noise was largely removed. Figure \ref{fig:lightcurve} shows the GJ~436 time series before and after the 2D spline correction. After the correction is applied, the precision of the light curve is 56\% above the photon limit. We note that the version of the {\it EPOXI} GJ~436 light curve presented in \cite{Ballard10} is very slightly different from the version used in this analysis. Because of the possibility of suppression of additional transits by the 2D spline correction method, we produced a version of the light curve which masked the points that occurred during transit times of the putative GJ~436c from contributing to the CCD sensitivity map. This set of additional masked points is 16 hours total in duration over the entire data set, consisting of 2 hours intervals centered at each of 10 candidate transit events-- two of which partially coincide with transits of GJ 436b and were already masked from the 2D spline correction. 

\subsection{Cold {\it Spitzer} 8 $\mu$m Observations}
Under {\it Spitzer} GO Program 50056 (PI: H. Knutson), GJ 436 was monitored for 70 hours continuously from 2008 July 12---15 by the {\it Spitzer Space Telescope} \citep{Werner04} with the IRAC subarray \citep{Fazio04} in the 8 $\mu$m channel, to obtain a full phase curve of the hot Neptune GJ~436b. This observing sequence consisted of 0.4 s integration exposures in 9195 blocks of 64 images each. These data were preflashed using the same technique as the HD 149026 observations (described in \citealt{Knutson09}). Preflashing effectively removes most of the ``detector ramp'' effect, which is characterized by an initial upward asymptote in the measured flux, followed by a gradual downward slope \citep{Charbonneau08,Knutson08,09aKnutson}. In this case we observed a region in M17 centered at RA 18$^{\mbox{h}}$20$^{\mbox{m}}$28$^{\mbox{s}}$ and Dec -16$^{\circ}$12'20'' with fluxes ranging between 3500-8000 MJy/Sr for 30 minutes prior to the start of the observations. This significantly reduced the amplitude of the detector ramp in these data, although there is still a small rise in flux of approximately 0.05\% visible in the first five hours of observations.  The data after this point have a flux variation of less than 0.05\% over the remaining 65 hours. The photometric extraction was performed by a similar method to the one described in detail in \cite{Knutson09}, which we summarize here. We determined the position of the star on the array by centroid, taking a position-weighted sum of the flux within a circular aperture with a radius of 5 pixels. We find that using this aperture size for the position determination returned the photometric lowest rms in the resulting time series. To perform background subtraction, we first created a median image from all the observations and identified four regions in the corners of the array where the light from the point spread function (PSF) was minimized, in order to minimize contamination by diffraction spikes in our background estimates. We created a histogram of the values of these pixels, and calculated the background from the mean of this distribution. We then performed aperture photometry with apertures ranging from 2.5 pixels in radius to 7.0 pixels in radius (the radius used for the aperture photometry affects the significance of the putative additional transit, which is explained in Section 3.2). We discarded points which lie more than 3$\sigma$ from the median value within each set of 64 images (in order to remove images affected by transient hot pixels). For the purposes of addressing the additional planet hypothesis, we were concerned with only a 5 hour portion of the light curve that fell in the latter half of the 70 hour phase curve, where we no longer observe any evidence for a detector ramp. Nonetheless, we fit and divided a quadratic function in time during the window of interest in order to remove any remaining trends in the data, which could be due to spots on the star, position-dependent aperture losses, and the phase variation from the planet itself.


\subsection{Warm {\it Spitzer} 4.5 $\mu$m Observations}
We observed GJ~436 in subarray mode, using a 0.1 s integration time in the 4.5 $\mu$m channel. These observations span 18 hours over UT 28 Jan 2010---29 Jan 2010. The observing sequence consisted of 7640 blocks of 64 images each. We experimented with two methods of locating the position of the star on the array. We first found a flux-weighted sum of the position within a circular aperture with a radius of 3 pixels. We also experimented with determining the position of the star by fitting a PSF within a 3.5 pixel aperture (we found similar results with larger apertures). Using the 100--times--oversampled PSF provided by the {\it Spitzer} Science Center for the 4.5 $\mu$m channel, we performed a $\chi^{2}$ minimization in which we allowed the X and Y positions of the PSF, a multiplicative brightness factor, and an additive sky background value to vary. We compared a histogram of the positions to get a sense for the precision of each measurement technique, first subtracting a running median calculated individually for each point from the nearest 20 points in time (to account for positional drift). We determined that these histograms had similar width with the two methods: $6.0\times10^{-3}$ pixels in Y with PSF-fitting versus $6.4\times10^{-3}$ pixels in Y with centroid. However, the precision of the final time series was not improved by using the PSF-fitting measured positions; rather, the precision was degraded by 20\%. We attribute this degradation to the decreased positional precision of the PSF positions: although their bulk scatter is less than the scatter from centroiding, the precision with which we measure an individual position is set by the resolution of the PSF at 0.01 pixels, rather than the 1$\sigma$ error from the scatter in the centroided positions of $6.4\times10^{-3}$ pixels. We considered the possibility of improving our positional accuracy by producing a more highly oversampled PSF with an interpolation of the {\it Spitzer} Science Center PSF, but found the computing time of PSF-fitting with such a large PSF to be prohibitively expensive.

 We then measured the stellar brightness in each image by performing aperture photometry on the Basic Calibrated Data (BCD) products across a range of apertures from 2.1 to 6 pixels. We calculate the sky background, which is almost negligable, from a 3$\sigma$-clipped mean of the pixels inside a ring of width 10 pixels from a radius of 7 pixels to 17 pixels. We found that the precision of the final time series was optimized at an aperture of 2.1 pixels in radius. 

\begin{figure}[h!]
\begin{center}
 \includegraphics[width=6in]{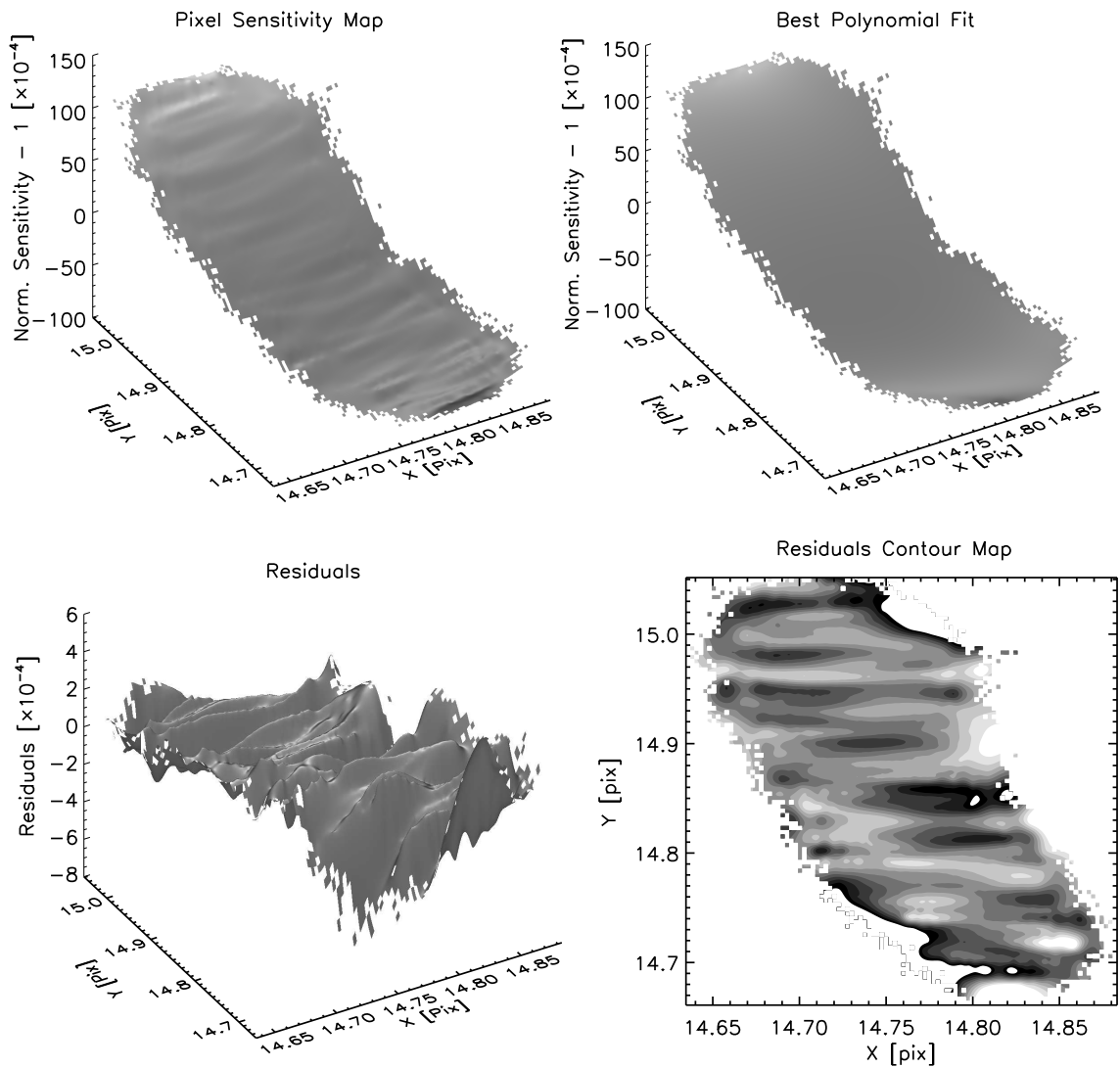} 
 \caption{\textit{Top panels:} At left, a three-dimensional view of the intrapixel sensitivity of the IRAC detector at 4.5 $\mu$m ($W(x_{i},y_{i})$, as computed in Equation \ref{eq:weight}). At right, the best fit third degree polynomial fit in X and Y. In both plots, The X axis shows the X position in pixels, the Y axis shows the Y position in pixels, and the Z axis show the fractional sensitivity. The jagged edges of the map are due to the low density of coverage in those areas; the star spent most of the time in the central region. \textit{Bottom panels:} At left, the residuals after the best-fit 2D polynomial is divided from the weighting function $W(x_{i},y_{i})$.  With large scale variations divided out, additional higher-frequency features are also visible. At right, a contour map of the same residuals.}
  \label{fig:pixelmap}
\end{center}
\end{figure}

We then corrected for the well-known intrapixel sensitivity variation observed in IRAC Channels 1 and 2 \citep{Reach05, Charbonneau05, Knutson08, Knutson09}. In lieu of fitting a polynomial in X and Y to the brightness variations, we instead implemented a point-by-point correction. We first binned the light curve into 20 second bins (approximately 145 points/bin). For each binned point, we evaluate a weighted sensivity function using all unbinned points (excluding those points which occurred inside the bin in question and outliers more than 3$\sigma$ from the mean of nearby points) given by:

\begin{equation}
W(x_{i},y_{i})= \frac{\displaystyle\sum_{j\neq i} exp{\left(-\frac{(x_{j}-x_{i})^{2}}{2\sigma_{x}^{2}}\right)} exp{\left(-\frac{(y_{j}-y_{i})^{2}}{2\sigma_{y}^{2}}\right)}\cdot f_{j}\cdot S(t_{j})}{\displaystyle\sum_{j\neq i} exp{\left(-\frac{(x_{j}-x_{i})^{2}}{2\sigma_{x}^{2}}\right)} exp{\left(-\frac{(y_{j}-y_{i})^{2}}{2\sigma_{y}^{2}}\right)}\cdot S(t_{j})},
  \label{eq:weight}
\end{equation}

where $f_{j}$ is the flux value of the $j$th observation, which is assigned a weight based on its distance in X and Y from the $i$th point being corrected. The function $S(t_{j})$ is a boxcar function in time, which is 1 for all points which are permitted to contribute to the sensitivity map, and 0 for all points which are not. The position of this function defines a ``mask'' interval during which time observations do not contribute to the interpixel sensitivity map. The purpose of this mask is to exclude points that do not reflect an accurate representation of the pixel sensitivity, such as those during transit; if we were to include these points in the creation of the intrapixel sensitivity map, we would both suppress the transit and introduce additional correlated noise outside of transit. The sensitivity function is then normalized by dividing by the Gaussian function multiplied by the boxcar $S(t_{j})$. We find the best results using $\sigma_{x}=0.017$ pixels and $\sigma_{y}=0.0043$ pixels (the width of the Gaussian weighting function is much smaller in Y because the dependence of the flux on the Y position is much stronger). We then divide the $i$th binned flux value by $W(x_{i},y_{i})$ to remove the effects of intrapixel sensitivity. Using a point-by-point sensitivity function, we do not need to assume a functional form for the intrapixel sensitivity; however, there is the very important caveat that flat-fielding the data by itself has the effect of suppressing the depths of additional transits, if they are present. We discuss how we avoid this effect in Section 3.3 with the use of the mask function. We show in Figure \ref{fig:pixelmap} three dimensional views of the intrapixel sensitivity map given by the weighting function $W(x_{i},y_{i})$. The large scale features of $W(x_{i},y_{i})$ are well approximated by polynomials in X and Y, per previous techniques, but we also find a smaller scale ``corrugation'' effect in the Y direction, where the sensitivity exhibits low-level sinusoidal-like variations with a separation of approximately 5/100ths of a pixel between peaks.

\begin{figure}[h!]
\begin{center}
 \includegraphics[width=6in]{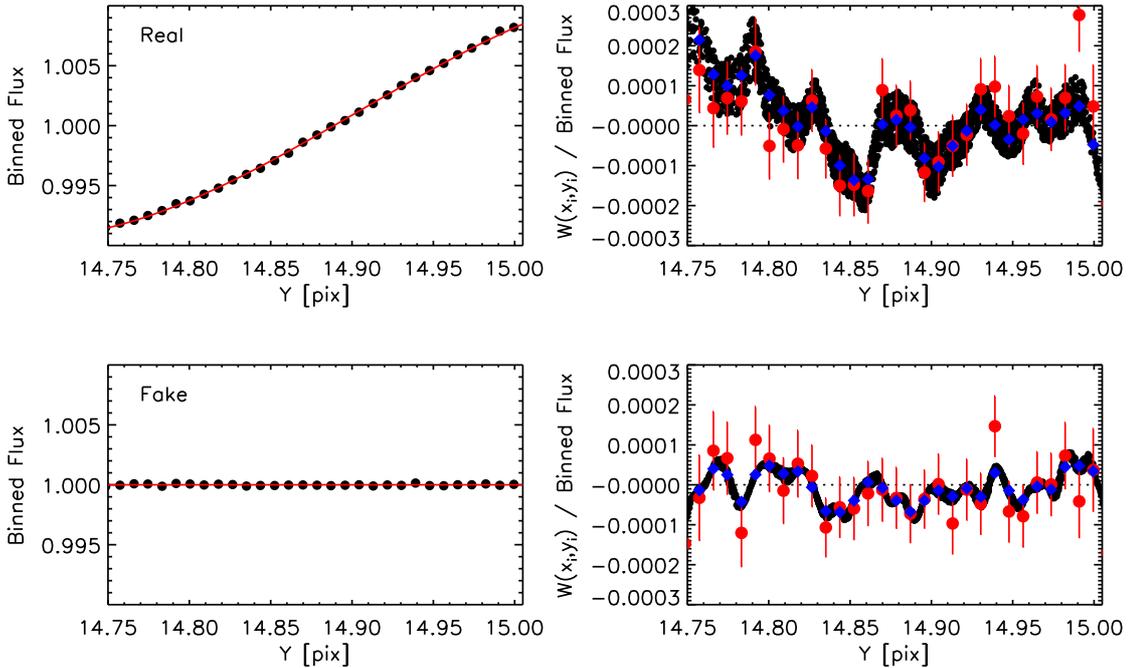} 
 \caption{\textit{Top panels:} At left, cross section of the flux as a function of Y position, holding X constant at 14.759 pixels (black points shown flux within 0.02 pixels of this X position, binned at 0.001 pixel intervals in Y; error bars are contained within the plotting symbol). Overplotted in red is the best-fit third degree polynomial. At right, the red points show the binned flux after the best-fit third degree polynomial is divided out. The black points overplotted shown the weighting function with best-fit third degree polynomial removed, and the blue points show the weighting function binned at the same intervals as the flux. \textit{Bottom panels:} At left, cross section of fake Gaussian data set over the same Y range-- no underlying structure is present. At right, the binned flux is shown in red, the weighting function is shown in black, and the binned weighting function is shown in blue.}
  \label{fig:pixelmap3}
\end{center}
\end{figure}

We investigated the authenticity of these features using several tests. For Gaussian noise, we expect the weighting function to exhibit only smoothed random noise, with features equal in size to the smoothing kernel, and so the corrugation features in Y should have a size scale near 9/1000ths of a pixel (because $\sigma_{y}$ is 0.0043 pixels). Futhermore, we should be able to predict the $\chi^{2}$ improvement by comparing the data to $W(x_{i},y_{i})$ as opposed to the null hypothesis. A Gaussian time series will be better fit by a smoothed version of itself than a flat line. The $\chi^{2}$ should improve by the number of smoothing kernels contained in the interval in question: for a time series with $N_{y}$ values ranging over 0.25 of a pixel, and a smoothing kernel of $\sigma_{y}$=0.0043, then the number of smoothing kernels (defining 3$\sigma$ from the center as the extent of the kernel) contained in the interval is approximately 10, and we should expect the $\chi^{2}$ of the fake time series to improve by 10 when compared to the weighting function instead of a flat line. Conversely, in order for authentic features attributable to the intrapixel sensitivity to be believable, the features must be significantly larger than the smoothing kernel, and the weighting function must provide a much better fit to the data than predicted from a smoothed version of the data itself. To test this hypothesis, we created a random Gaussian data set sampled at the same X and Y positions on the detector, with a standard deviation equal to that of the actual time series. We then created a sensitivity function using the same kernel in X and Y as for the actual observations, and compared the results; this comparison is shown in Figure \ref{fig:pixelmap3}. 

We find that the weighting function from the random data set looks as we expected it to look, with corrugation features near the size of the FWHM of the smoothing kernel. The improvement in $\chi^{2}$, using a flat line with value 0 as compared to the binned weighting function, is 11.4 for the fake data set, very close to the predicted improvement of 10. For comparison, the improvement in $\chi^{2}$ for the real data, between a flat line and the weighting function, is 24.3, and the amplitude of the features are larger. Furthermore, the peak-to-peak size scale of features is 0.05 pixels, more than 6 times the FWHM of the smoothing kernel, which argues against their being smoothing artifacts. The difference in width of the weighting function between the real and fake data sets (the thickness of the black points in the vertical direction) can be attributed to the weak X dependence of the pixel sensitivity. In the lowest right hand panel of Figure \ref{fig:pixelmap}, there is evidence for pixel sensitivity variation in the X direction over a size scale comparable to the width of the 0.04 pixel cross section used to generate Figure \ref{fig:pixelmap3}. For these reasons, we are confident that the high-frequency features, with a period of 0.05 pixels, in the real weighting function are authentic. The amplitude of this effect is 100 ppm, 40\% of the amplitude of a 0.75 $R_{\oplus}$ transit in front of a 0.438 $R_{\odot}$ star such as GJ~436 \citep{Ballard10}. Therefore, removing this correlation with position was crucial to our ability to rule out the putative transit of depth 250 ppm.

With this photometric reduction procedure, we achieve a precision of 0.0053 per 0.1 s exposure on the unbinned time series. Compared to the photon noise-limited precision of 0.0042, we are 26\% above the photon noise limit, although the presence of remaining correlated noise means that the scatter with larger bin sizes deviates more from the ideal Gaussian limit. We achieve a sensitivity of 71 ppm per 20-minute bin, compared to the shot-noise limit of 41 ppm at that bin size. 

For comparison, we also perform a reduction using a polynomial fit in X and Y to remove the intrapixel sensitivity variations in flux \citep{Reach05, Charbonneau05, Knutson08, Knutson09}. We express the measured flux $f'$ in terms of the incident flux $f$ and the X and Y position of the star on the detector with the following expression:

\begin{equation}
f'=f\left(b_{1}+b_{2}(x-\bar{x})+b_{3}(x-\bar{x})^{2}+b_{4}(y-\bar{y})+b_{5}(y-\bar{y})^{2})+b_{6}(y-\bar{y})^{3})\right)
\end{equation}

We find that the precision we achieve using a polynomial intrapixel sensitivity function, for the same bin size of 20 minutes, is 230 ppm, as compared to a precision of 71 ppm using the weighted sensitivity map $W(x_{i},y_{i})$. If we divide the time series into 3 portions of 5 hours duration, and fit separate polynomial coefficients for each portion, we achieve a precision of 91 ppm for a bin size of 20 minutes-- still 1.3 times times larger than the precision using $W(x_{i},y_{i})$. However, although we can improve the overall precision of the time series by fitting the polynomial coefficients independently for increasingly short durations, we are never able to reliably recover transits of a 0.75 $R_{\oplus}$ planet in a time series reduced with a polynomial sensitivity function. We discuss this analysis in Section 3.3.

\section{Search for Photometric Evidence}

\subsection{The Suggestion from {\it EPOXI} }

In \cite{Ballard10}, we conducted a search for additional transiting planets in the {\it EPOXI} light curve of GJ~436. In that work, we demonstrate our sensitivity to additional transits by injecting light curves corresponding to additional planets with varying planetary radius, period, and phase, and then attempting to recover them by maximizing the $\chi^{2}$ goodness of fit. We found, when we carefully accounted for the signal suppression introduced by reducing the data with the 2D spatial spline method, we were sensitive to Earth-sized planets with good ($\ge$50\%) probability for periods less than only 0.5 days. However, we discovered weak evidence for an additional transiting planet, which fell well below the criterion we established for a detection. In \cite{Ballard10}, we empirically established the criterion for detection that used the improvement of $\chi^{2}$ corresponding to the best fit transit signal compared to the $\chi^{2}$ of the null hypothesis: we could reliably recover the correct period of any signal which produced an improvement of $\Delta\chi^{2}\ge$250. The transit signal corresponding to a 0.75 $R_{\oplus}$ size planet is well below this threshold. However, we find that the largest deviations occurred with a regular period near the 4:5 resonance with the Neptune-sized planet. Five of these events produced a combined improvement to the $\chi^{2}$ of 140. However, these 5 comprise only half of the transits we would expect to see with this ephemeris: Of these remaining 5 events, two coincide with transits of GJ~436b, one occurs during a gap in the phase coverage, and two show no evidence of a transit. If we include all predicted candidate transit events in our $\chi^{2}$ calculation (including the two events that coincide with a transit of GJ~436b), the null hypothesis gives a better solution than any transit model. However, if points that coincide closely in time with transits of the GJ~436b are excluded from the calculation, the improvement over the null $\chi^{2}$ is 70. There are two motivations to exclude these in-transit points: First, we fit a slope with time to the points immediately outside of the transit of GJ~436b (from 3 minutes to 30 minutes before the start of transit and after the end of transit) and divide this slope in order to normalize each transit before we fit for the system parameters (see \citealt{Ballard10}), and this procedure may suppress other signals. Second, there is also the small possibility of an occultation of GJ~436c by GJ~436b.

\begin{figure}[h!]
\begin{center}
 \includegraphics[width=3in]{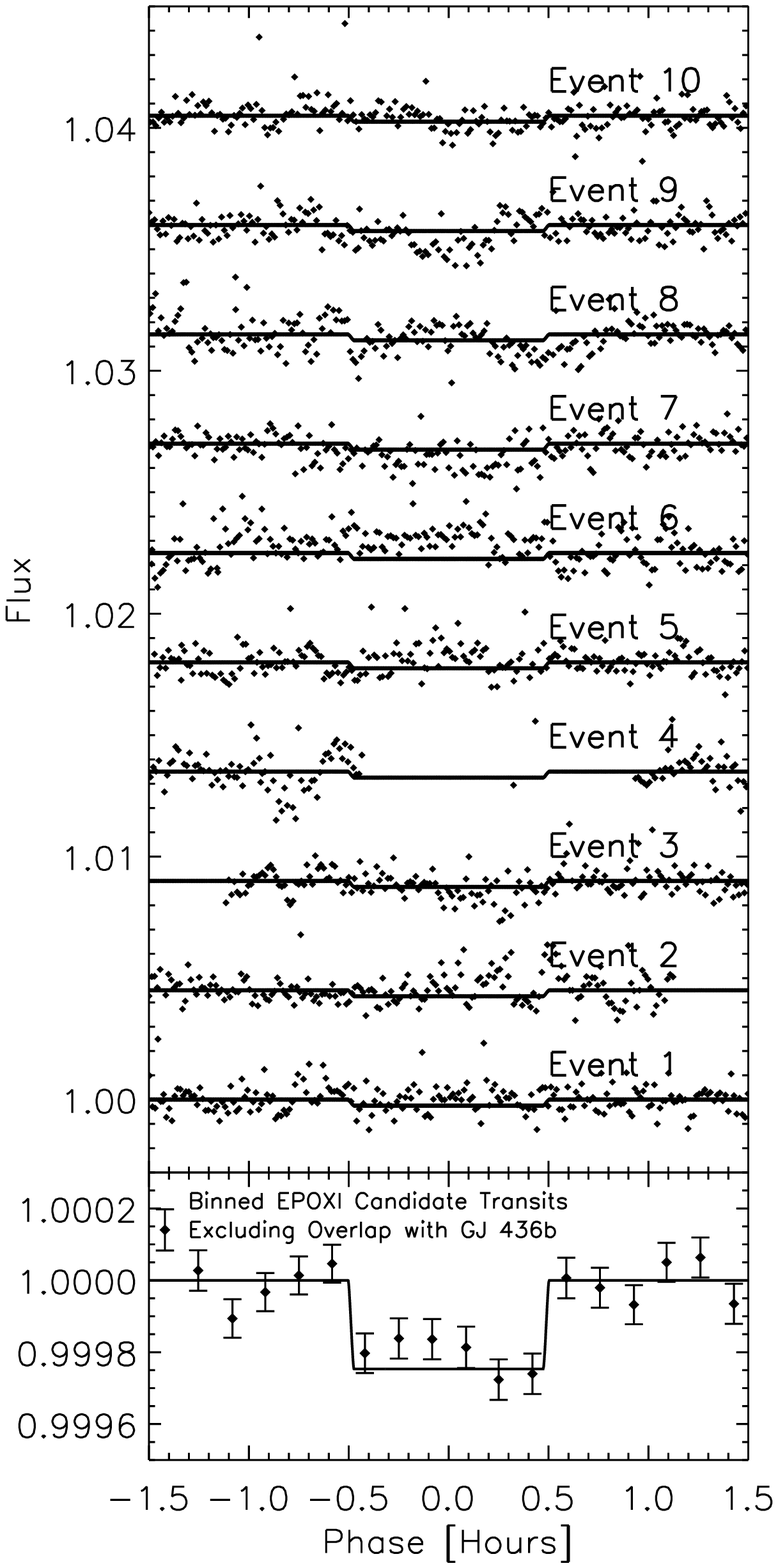} 
 \caption{\textit{Top panel:} {\it EPOXI} time series of GJ~436, after dividing out model of transits of GJ~436b, during times of transit of the putative GJ~436c discussed in text. The significance of the light curve shown overplotted is far below the criterion of a confident detection \citep{Ballard10}. Model light curves \citep{Mandel02} with the predicted transit depth and ephemeris are shown overplotted with the solid black line; the events with positive transit depth are labeled Events 3, 7, 8, 9, and 10. A gap in the phase coverage occurs during predicted Event 4, Events 2 and 5 show negatives deviations (an increase in brightness, as opposed to a decrement), and Events 1 and 6 overlap in time with transits of GJ~436b. \textit{Bottom panel:} All events except those that overlap in time with a GJ~436b transit (labeled 1 and 6 above), binned in 10 minute intervals.}
  \label{fig:epoxi}
\end{center}
\end{figure}

We observed comparable $\Delta\chi^{2}$ improvement (within 2$\sigma$) for periods ranging between 2.1074 days (0.797 the period of GJ~436b) and 2.1145 days (0.800 the period of GJ~436b). Figure \ref{fig:epoxi} shows the 10 events separately. We also plot a transit model corresponding to a planet with radius 0.75 $R_{\oplus}$ and period 2.1076 days (this period was selected by combining the {\it EPOXI} and {\it Spitzer} 8 $\mu$m observations described in the next section).

\subsection{Corroboration by {\it Spitzer} at 8 $\mu$m}
\begin{figure}[h!]
\begin{center}
 \includegraphics[width=6in]{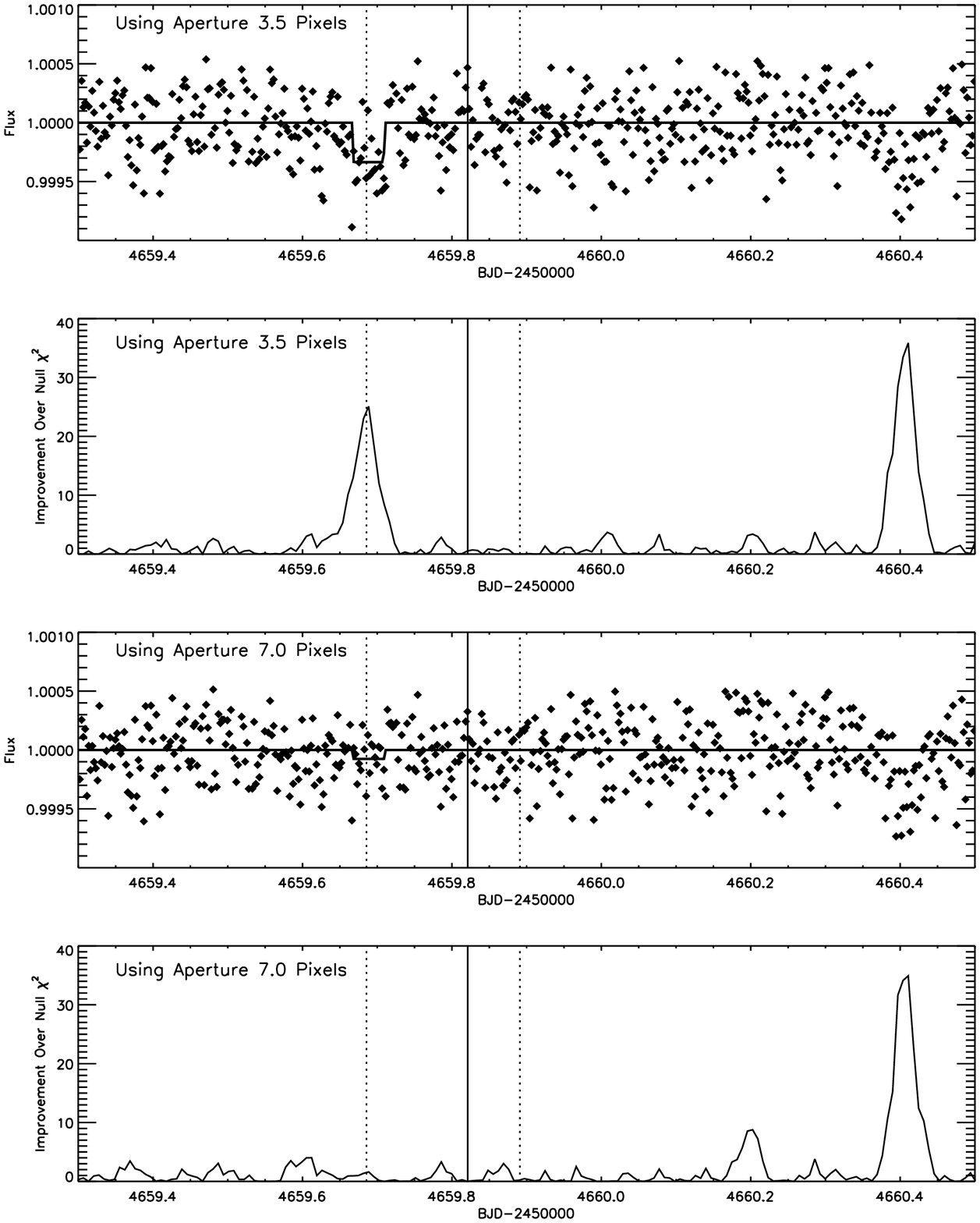} 
 \caption{\textit{Top two panels:} Portion of {\it Spitzer} 8 $\mu$m phase, photometry performed with aperture radius of 3.5 pixels, with solid line denoting the putative transit time predicted from {\it EPOXI} and the dashed lines indicating the 2$\sigma$ confidence interval. The best-fit transit light curve solution is shown overplotted in red, with a depth of 1 $R_{\oplus}$ at this aperture, and the secondary eclipse of GJ~436b is evident at a time 17.5 hours after the candidate transit. The second panel shows the improvement of the best-fit transit light curve solution $\chi^{2}$ at each time compared to the null $\chi^{2}$. \textit{Bottom two panels:} The same portion of the 8 $\mu$m phase curve, with an aperture radius of 7.0 pixels. The last panel shows the improvement of the best-fit transit light curve solution as compared to the null.}
  \label{fig:spitzerch4}
\end{center}
\end{figure}

The constraints on the ephemeris of the putative GJ~436c from the {\it EPOXI} data alone meant that the accuracy with which we could predict the times of transits 1.5 years out from the {\it EPOXI} observations was poor. However, the extant {\it Spitzer} 8 $\mu$m phase curve was gathered only two months after the {\it EPOXI} observations took place, such that our accuracy on the predicted time was 5 hours (defined by the duration of the 2$\sigma$ confidence interval from {\it EPOXI}). We performed a boxcar search of this light curve, allowing both the time of transit and the depth of transit to vary, since the {\it EPOXI} observations provided only weak constraints on the planetary radius. We discovered a transit-like signal in these data within the time window predicted from {\it EPOXI}, but the signal was present (that is, produced a $\Delta\chi^{2}$ improvement larger than any other feature during the time window of interest) only when apertures smaller than 4 pixels in radius were used to perform the photometry. The top panel of Figure \ref{fig:spitzerch4} shows a portion of the 8 $\mu$m light curve, with 3.5 pixel aperture used to extract the photometry. The solid black curve shows the best-fit transit light curve solution, and the event at a time of BJD 2454660.4 is the secondary eclipse of the hot Neptune GJ~436b. The second panel shows the significance of the best $\chi^{2}$ as a function of time. The bottom two panels show the results using a 7.0 pixel aperture to extract the photometry. While the significance of the secondary eclipse remains constant, the significance of the putative additional transit depends on aperture size. 

The dependence of the putative transit signal on the size of the aperture argued against the planet hypothesis: rather, a signal that disappears at larger radii is more likely due to position-dependent flux losses. We concluded that the {\it Spitzer} 8 $\mu$m observations neither definitely confirmed nor refuted the planet hypothesis, so we gathered additional {\it Spitzer} observations at 4.5 $\mu$m, where we could obtain a higher precision light curve, to definitely resolve the question. The single candidate transit from {\it Spitzer} greatly decreased the possible parameter space of the planet's ephemeris, so that we were able to predict a transit to occur 1.5 years after the {\it EPOXI} observations within an 15 hour window. We also predicted the radius of the putative planet, from a $\chi^{2}$ minimization of the combined {\it EPOXI} and {\it Spitzer} 8 $\mu$m observations, to be 0.75 $R_{\oplus}$.

\subsection{The Death Knell from {\it Spitzer} at 4.5 $\mu$m}
The method we use to correct the 4.5 $\mu$m data has the possible effect of suppressing transit signals, since the point-by-point correction relies critically on the assumption that the flux variations are due {\it only} to the pixel sensitivity variations, and that the stellar flux is constant. If a transit occurs during the observations, then the derived value of the pixel sensitivity $W(x_{i},y_{i})$ in the location of the detector where the transiting points occur will be lower by a fraction that depends on the pixel sampling. If the points in transit comprise 10\% of the observations in that location on the detector, then the value for the pixel sensitivity in that location will be low by 10\%. This will have the effect of suppressing the transit depth by 10\%, since we will incorrectly attribute a fraction of the decrement at that time to the pixel performance, rather than an astrophysical variation. Therefore, to correctly recover the true transit signal, we must iteratively identify and then mask the points which occur in transit. This presents a challenge since we do not know exactly when the transit should occur. We tested the procedure of simply masking points that occur within a transit duration of each point being corrected (so that the sensitivity function for each point is calculated using points more than half a transit duration removed in time), but found that this did not produce the highest quality time series. Although this masking procedure prevents suppression of the transit signal when the mask is corrected located over the transit, the transiting points are allowed to contribute to the sensitivity function at other times, which introduces correlated noise to the remainder of the time series. Although the depth of the transit signal is preserved, its significance relative to other features in the time series is dimished, reducing our ability to correctly identify it. We therefore concluded that the correct procedure is to locate the position of the transit in time, and mask the set of transiting points for each individual flux correction. The challenge is to locate the true position of the transit in order to correctly place the mask. We addressed this question by producing $n$ time series, where $n$ is the number of tested mask positions (in this case, we tested mask positions in 30 minute intervals over a roughly 15 hour time series, resulting in 33 mask positions). We hypothesize that when the mask correctly coincides with a transit, both the depth of the transit and its significance relative to the next most significant feature in the time series should increase, thus enabling us to identify the time of transit.   

\begin{figure}[h!]
\begin{center}
 \includegraphics[width=6in]{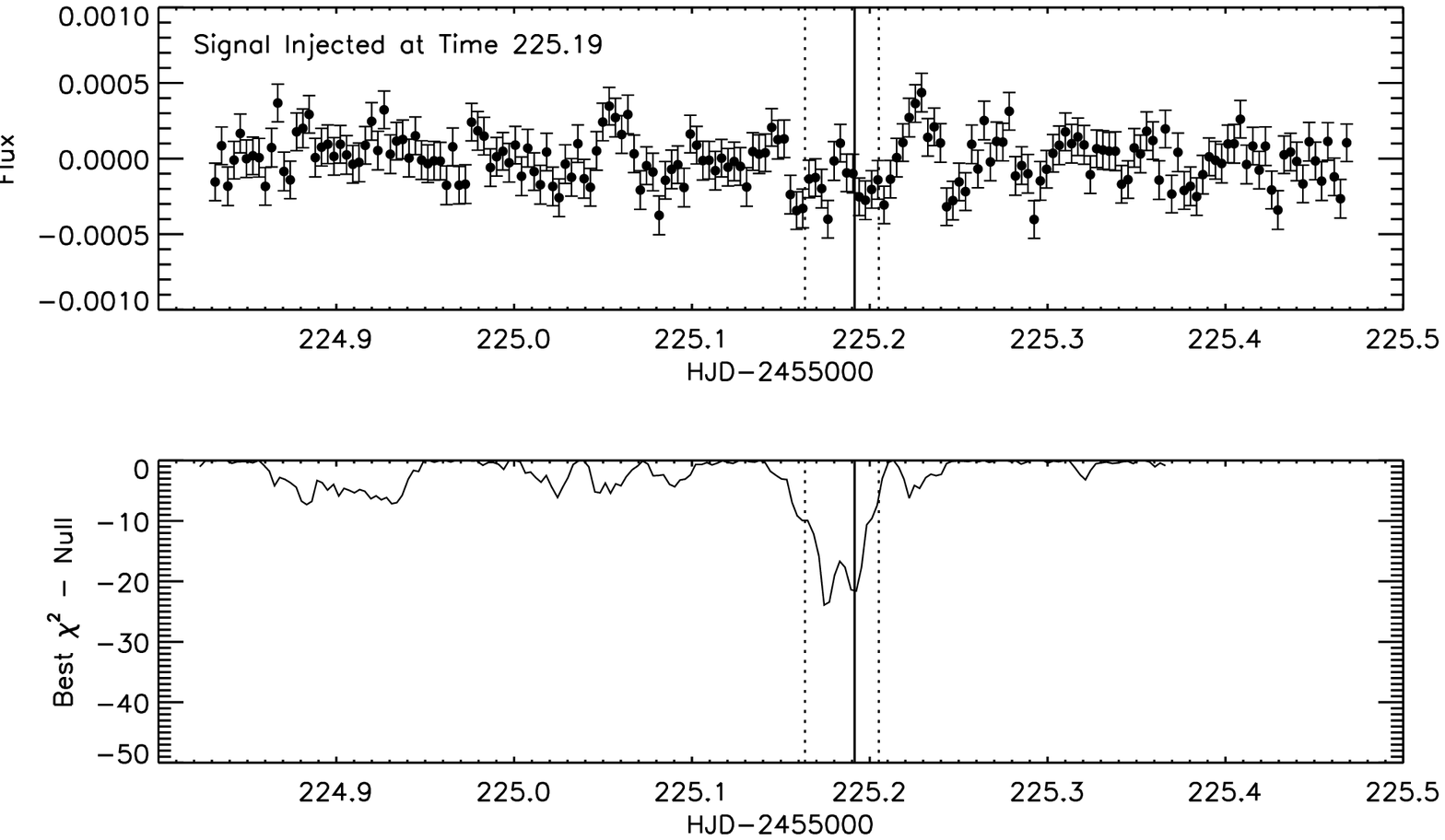} 
 \caption{\textit{Top panel:} {\it Spitzer} 4.5 $\mu$m light curve, with 0.75 $R_{\oplus}$ planet transit injected at time shown by the solid line. The location of the time window during which points are masked from the weighting function is shown by the dashed lines. \textit{Bottom panel:} The improvement of the best-fit transit light curve solution $\chi^{2}$ at each time compared to the null $\chi^{2}$. In this case, the significance doubles when the mask is correctly placed over the transit.}
  \label{fig:spitzerch2_inject}
\end{center}
\end{figure}

In order to establish that we could reliably detect the 0.75 $R_{\oplus}$ radius planet, we injected a transit of this size into the {\it Spitzer} 4.5 $\mu$m observations. We attempted to blindly recover the injected transit time by varying the mask position (which was always one hour in duration) by 30 minute intervals, producing a separate time series for each mask position. Then, for each of these $n$ time series, we evaluated the significance of a boxcar light curve function with the predicted transit depth and duration, allowing only the time of the transit to vary. We identified the time of transit from the time of most significant improvement to the $\chi^{2}$ from the boxcar search among all the time series, which indeed occurred when the mask was correctly located over the injected transit. We then repeated this procedure, shifting the injected transit time in hour increments, to ensure that we could detect the transit at any time during the observations. Figure \ref{fig:spitzerch2_inject} shows the 4.5 $\mu$m light curve, with 0.75 $R_{\oplus}$ transit signal injected at a location denoted by the solid line. The dashed lines mark the beginning and end of the window in time during which data are masked from the sensitivity function calculation. In this case, the significance of the detection, using the improvement over the null $\chi^{2}$, increases by a factor of two when the mask is correctly located over the transit, which enabled us to blindly locate the time of the injected transit.


\begin{figure}[h!]
\begin{center}
 \includegraphics[width=6in]{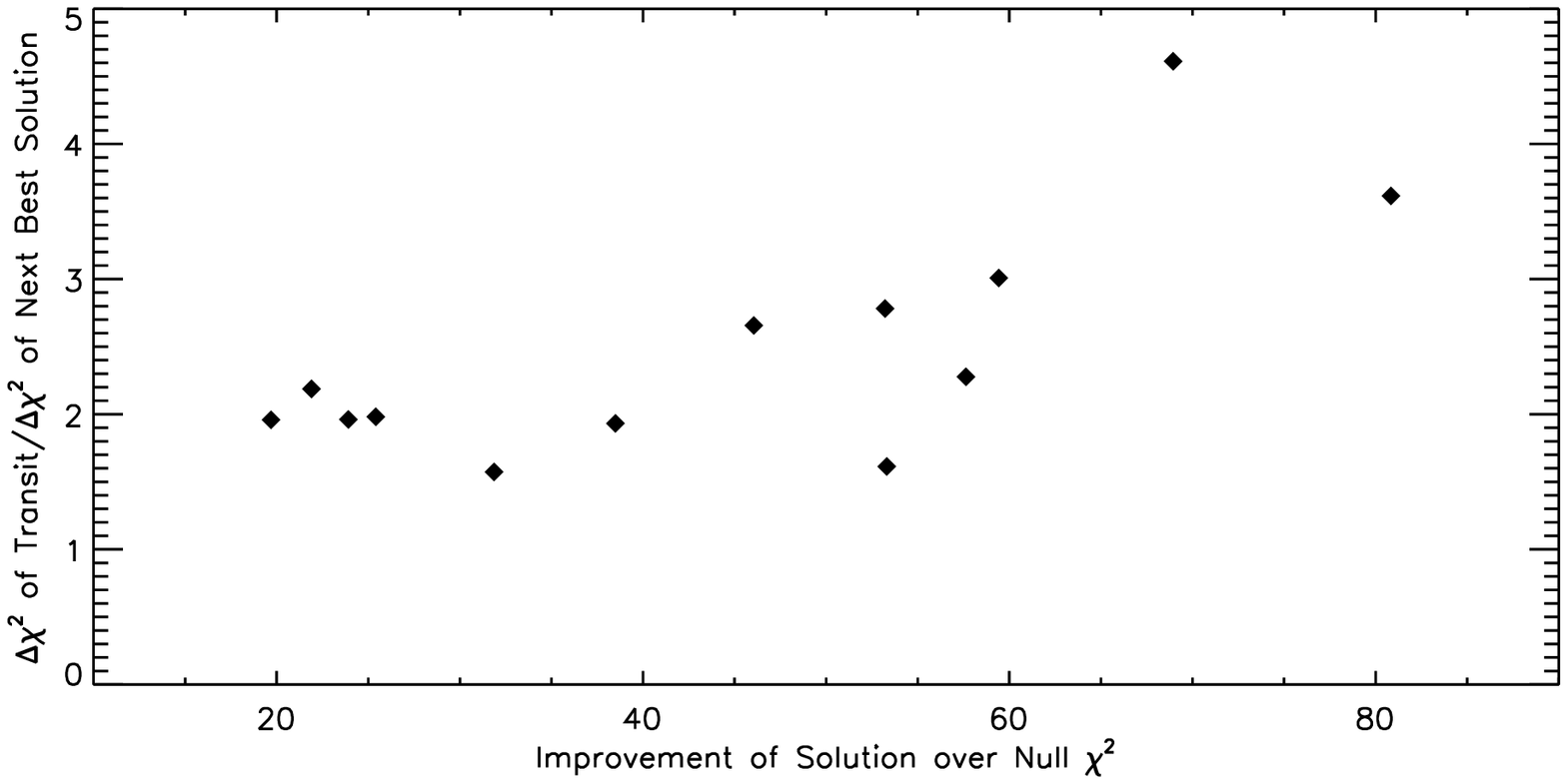} 
 \caption{Recovery statistics of 0.75 $R_{\oplus}$ transit signals injected into {\it Spitzer} 4.5 $\mu$m observations. All transits times within the window allowed by the {\it EPOXI} and {\it Spitzer} 8 $\mu$m were successfully recovered. The X axis gives the absolute significance of the recovery, and the Y axis gives the ratio of the significance of the transit to the significance of the next best solution.}
  \label{fig:spitzerch2_recovstats}
\end{center}
\end{figure}

We assess our sensitivity to 0.75 $R_{\oplus}$ planet transits by noting both the absolute significance of the detection, and the ratio of the detection significance to the significance of the next best solution, for all injected signals. We find that, for all transits occurring after a time of BJD 2455224.9 (corresponding to a period of 2.1071 days, which is safely before any ephemeris consistent with the solution from {\it EPOXI}), we recover the correct transit time with significance $\Delta\chi^{2}\gtrsim20$, and in all cases the significance of the transit signal (once the mask is centered over the transit time, so the suppression is minimized) is at least 60\% higher than the the significance of the next highest solution. Figure \ref{fig:spitzerch2_recovstats} shows the recovery statistics for all injected transit times, starting at BJD 2455224.90 and ending at BJD 245525.44, in increments of 1 hour. 


\begin{figure}[h!]
\begin{center}
 \includegraphics[width=6in]{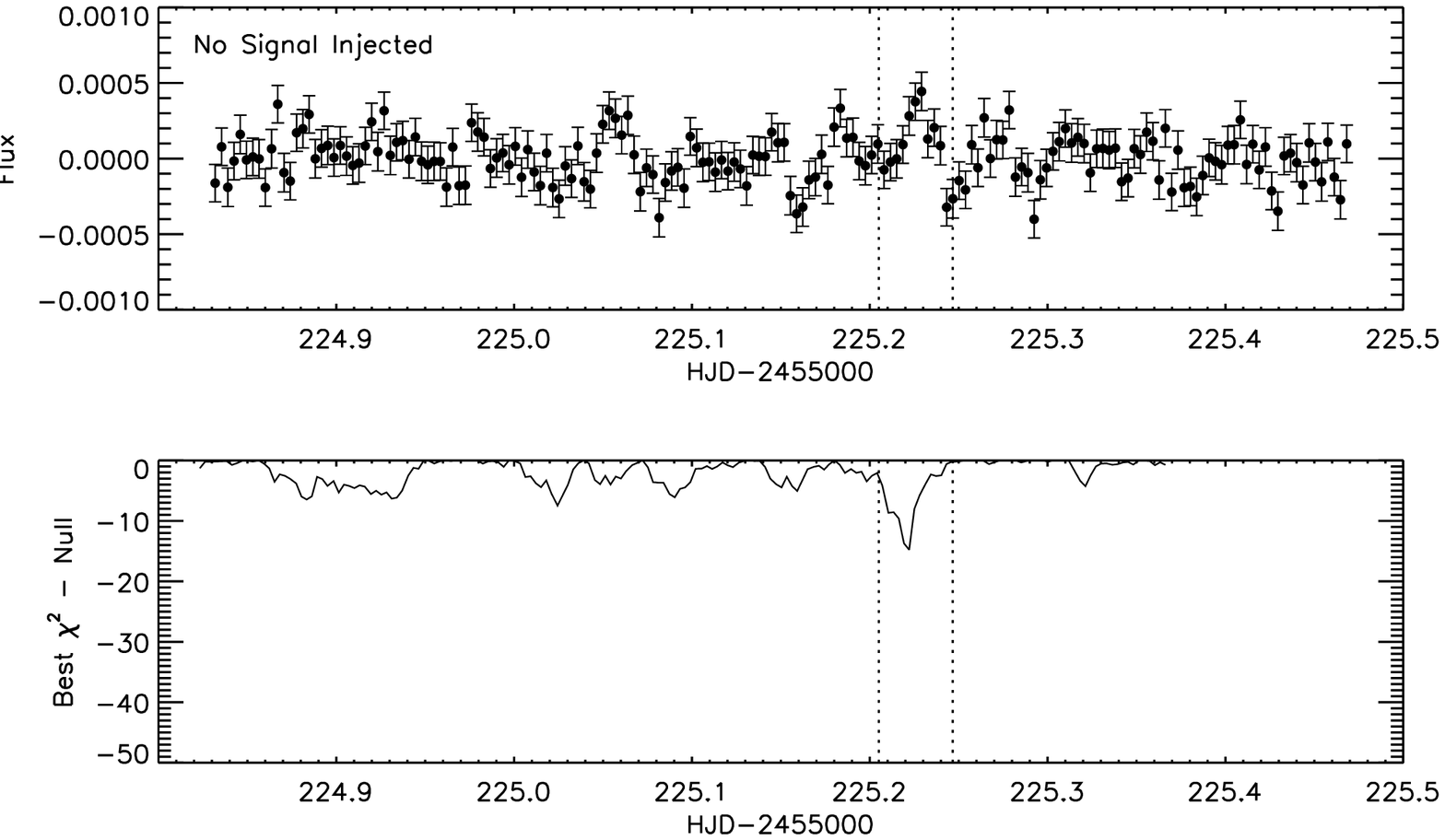} 
 \caption{\textit{Top panel:} {\it Spitzer} 4.5 $\mu$m light curve. The location of the time window during which points are masked from the weighting function is shown by the dashed lines; using this window, we find the most significant ``transit'' signal, which is actually an anti-transit. \textit{Bottom panel:} The improvement of the best-fit transit light curve solution $\chi^{2}$ at each time compared to the null $\chi^{2}$.}
  \label{fig:spitzerch2_noinject}
\end{center}
\end{figure}

We note also that the average detection significance is a $\Delta\chi^{2}$ of 45, as compared to a predicted significance of 37 (using the scatter of 71 ppm for 20 minutes bins compared to the transit depth of 250 ppm, and assuming an hour-long transit). Having demonstrated our sensitivity to transits as small as the putative GJ~436c, we then repeated the above analysis on the actual time series---generating different versions of the time series for each mask position, while keeping the duration of the mask constant at one hour. The solution with the best improvement over the null hypothesis is shown in Figure \ref{fig:spitzerch2_noinject}, with the beginning and end of the interval during which data are masked from the weighting function shown by dashed lines. The best solution is actually an anti-transit in this case; when the mask is located over these points, the solution with highest significance gives an improvement over the null $\chi^{2}$ of 15. The most significant solution with a transit decrement solution has a significance of 7. We do not find any signal with the significance at which we detected injected signals of the expected depth. 

We repeated this analysis on a version of the time series which was reduced using a polynomial fit to the intrapixel sensitivity variation. The standard deviation of the time series, as discussed in Section 2.3, is 30\% higher than the standard deviation of the time series reduced using the weighted sensitivity function, even after we fit coefficients independently to 5-hour pieces of the time series. We find that we are able to recover the correct injected transit time only 20\% of the time when the time series is reduced using a polynomial fit to the sensitivity function.

\section{Discussion}
We conclude that the putative transiting sub-Earth-sized GJ~436c planet, which was suggested by our {\it EPOXI} and {\it Spitzer} 8 $\mu$m data, can be conclusively ruled out by our {\it Spitzer} 4.5 $\mu$m data. The periodicity of the candidate transit events within the {\it EPOXI} and {\it Spitzer} 8 $\mu$m data sets, coupled with the proximity of the hypothesized period to a resonance with the period of the known hot Neptune GJ~436b, initially merited further investigation, but the lack of a transit in the {\it Spitzer} 4.5 $\mu$m observations proves definitively that the candidate transit signals are not authentic. 

Motivated by the intriguing eccentricity of GJ~436b, the observational campaigns to find the putative additional planet responsible have resulted in very sensitive upper limits to GJ~436c. The radial velocity analysis presented by \cite{Bean08b} ruled out perturbers greater than 8 $M_{\oplus}$ at periods less than about 11 days (semi-major axes less than 0.075 AU) with high confidence. From the {\it EPOXI} search for additional transits of this putative planet, we ruled out rocky transiting bodies down to 9.6 $M_{\oplus}$ with periods less than 8.5 days with 95\% confidence in the GJ~436 system \citep{Ballard10}, in addition to definitively ruling out the 0.75 $R_{\oplus}$ planet suggested by the combined {\it EPOXI} and 8 $\mu$m {\it Spitzer} photometry. Furthermore, the possibility of a close-in resonant companion in 2:1 or 3:1 resonance with GJ~436b is strongly disfavored by transit timing measurements \citep{Pont09}. \cite{Batygin09} compiled a list of possible dynamically stable secular perturbers which are consistent with the transit times, radial velocities, and observed eccentricity of GJ~436b, which are observationally tractable. In light of the sensitive upper limits to this perturbing companion, the resolution to the eccentricity of GJ~436b may instead be a higher tidal dissipation factor for the hot Neptune---such a $Q$ would need to be 1--2 orders of magnitude larger than that measured for Neptune in our solar system \citep{Batygin09, Banfield92}. A value of $10^{6.3}$ for GJ~436b for $Q/k_{2}$ (where the Love number $k_{2}$ is typically near 0.5 for Solar System gas giants; \citealt{Bursa92}) proposed by \cite{Jackson08} could explain the eccentricity of GJ~436b without requiring the presence of an additional planet.

We demonstrate that the precision we obtain with Warm {\it Spitzer} observations at 4.5 $\mu$m is sufficient to detect a sub-Earth-sized planet around GJ~436. We find that the use of a polynomial to correct for the intrapixel sensitivity variation is insufficient to detect the putative 0.75 $R_{\oplus}$ planet. It was therefore necessary to correct for the variation with a point-by-point weighted sensitivity map in order to conclusively rule out the existence of the planet: both the rms precision and our ability to recover injected transit signals are enhanced with this correction method. We hope the methods outlined here to obtain this precision will be a useful guide for the reduction of future Warm {\it Spitzer} data sets of {\slshape Kepler} targets, some of which almost certainly {\it will} contain transits of Earth-sized planets. 

\bibliographystyle{apj} 

\end{document}